\documentclass[10pt]{article} 
\usepackage{graphicx}
\usepackage{hyperref}
\usepackage{amssymb}
\usepackage{aas_macros}
\usepackage{authblk}
\usepackage{color}
\newcommand{\mau}[1]{{\leavevmode\color{black}#1}}
\title{Search for neutrinos from TXS 0506+056 with the ANTARES telescope}
\author[1]{A.~Albert}
\author[2]{M.~Andr\'e}
\author[3]{M.~Anghinolfi}
\author[4]{G.~Anton}
\author[5]{M.~Ardid}
\author[6]{J.-J.~Aubert}
\author[7]{J.~Aublin}
\author[7]{T.~Avgitas}
\author[7]{B.~Baret}
\author[8]{J.~Barrios-Mart\'{\i}}
% ...

\affil[1]{\scriptsize{Universit\'e de Strasbourg, CNRS,  1 UMR 7178, F-67000 Strasbourg, France}}
\affil[2]{\scriptsize{Technical University of Catalonia, Laboratory of Applied Bioacoustics, Rambla Exposici\'o, 08800 Vilanova i la Geltr\'u, Barcelona, Spain}}
\affil[3]{\scriptsize{INFN - Sezione di 3, Via Dodecaneso 33, 16146 3, Italy}}
\affil[4]{\scriptsize{Friedrich-Alexander-Universit\"at 4-N\"urnberg, 4 Centre for Astroparticle Physics, Erwin-Rommel-Str. 1, 91058 4, Germany}}
\affil[5]{\scriptsize{Institut d'Investigaci\'o per a la Gesti\'o Integrada de les Zones Costaneres (IGIC) - Universitat Polit\`ecnica de Val\`encia. C/  Paranimf 1, 46730 Gandia, Spain}}
\affil[6]{\scriptsize{Aix Marseille Univ, CNRS/IN2P3, 6, Marseille, France}}
\affil[7]{\scriptsize{7, Univ Paris Diderot, CNRS/IN2P3, CEA/Irfu, Obs de Paris, Sorbonne Paris Cit\'e, France}}
\affil[8]{\scriptsize{8 - Instituto de F\'isica Corpuscular (CSIC - Universitat de Val\`encia) c/ Catedr\'atico Jos\'e Beltr\'an, 2 E-46980 Paterna, Valencia, Spain}}
% ...

\begin{document} 
\maketitle 
\begin{abstract}
The results of three different searches for neutrino candidates, associated with the IceCube-170922A event or from the direction of TXS 0506+056, by the ANTARES neutrino telescope are presented.
The first search refers to the online follow-up of the IceCube alert; the second is based on the standard time-integrated method employed by the Collaboration to search for point-like neutrino sources; the third uses the information from the IceCube time-dependent analysis reporting a bursting activity centered on December 13, 2014, as input for an ANTARES time-dependent analysis.
The online follow-up and the time-dependent analysis yield no events related to the source. The time-integrated study performed over a period from 2007 to 2017 fits 1.03 signal events, which corresponds to a p-value of 3.4\% (not considering trial factors). Only for two other astrophysical objects in our candidate list, a smaller p-value had been found. When considering that 107 sources have been investigated, the post-trial p-value for TXS 0506+056 corresponds to 87\%. 
\end{abstract}
%

\section{Introduction}

High-energy (HE) neutrinos are produced through the decay of charged mesons, and are associated with $\gamma$-rays resulting from the decay of neutral mesons. These mesons are previously produced in the interactions of protons or nuclei with ambient matter or radiation. Thus, the observation of neutrinos associated with known sources of $\gamma$-rays and/or electromagnetic radiation provides the identification of cosmic objects where hadrons are accelerated.

The IceCube Collaboration has reported a significant excess of a diffuse flux of HE astrophysical neutrinos over the atmospheric background \cite{IC1,IC2}. However, no individual HE neutrino source have been identified so far.
A recent HE neutrino detected by the IceCube experiment was connected with observations in $\gamma$-rays and at other wavelengths of the electromagnetic spectrum from blazar TXS 0506+056 \cite{ICgEM}. This may indicate that this blazar (a BL Lac object, at redshift $z= 0.3365 \pm0.0010$ \cite{paiano}) is an individually identifiable source of HE neutrinos. Two independent analyses (one \textit{time-integrated} and one \textit{time-dependent}) of IceCube data searching for neutrino emission at the position of the blazar confirmed an excess of events at the level of 3.5$\sigma$ \cite{ICalo}.

The ANTARES \cite{Ageron11} telescope is a deep-sea Cherenkov neutrino detector, located 40 km off shore from Toulon, France, in the Mediterranean Sea.
The telescope aims primarily at the detection of neutrino-induced 
muons that cause the emission of Cherenkov light in the detector (\textit{track-like} events). Charged current interactions induced by electron neutrinos (and, possibly, by tau neutrinos of cosmic origin) or neutral current interactions of all neutrino flavors can be reconstructed as \textit{shower-like} events \cite{albertSho}. 
Due to its location, the ANTARES detector mainly observes the Southern sky ($2\pi$ sr at any time).
Events arising from sky positions in the declination band $-90^\circ \le \delta \le -48^\circ $ are always visible as upgoing. 
Neutrino-induced events in the declination band $-48^\circ \le \delta \le +48^\circ $ are visible as upgoing with a fraction of time decreasing from 100\% down to 0\%.  

In this document, the results of three different searches for neutrino candidates associated with the IceCube-170922A event or from the direction of TXS 0506+056 by the ANTARES neutrino telescope are presented.
The first search refers to the online follow-up associated with IceCube-170922A (Sect. \ref{sec:an1}).
The second is based on the standard likelihood method employed by the Collaboration to search for point-like neutrino sources (Sect. \ref{sec:an2}).
The third uses the information from the time-dependent analysis performed by the IceCube Collaboration \cite{ICalo}, which reports a bursting activity centered on December 13, 2014, as input for an ANTARES time-dependent analysis.
Conclusions are reported in Sect. \ref{sec:an4}.

\section{Online searches for neutrinos in ANTARES associated to IceCube-170922A EHE
\label{sec:an1}}

Following the IceCube observation of a HE neutrino candidate event, IceCube-170922A, which occurred at $T_0$ = 17/09/22, 20:54:30.43 UT (in the following referred to as IC170922A), the ANTARES Collaboration performed an online follow-up analysis to look for additional neutrinos from the reported event direction. 

The IceCube event was identified by the Extremely High Energy (EHE) track event selection and reported with a GCN circular \cite{ICGCN}. This EHE event had a high probability of being of astrophysical origin and the IceCube Collaboration encouraged follow-up to help identifying a possible astrophysical source. 
The reported position (J2000, with 90\% point-spread function (PSF) containment) was RA=$77.43^\circ \ _{-0.80^\circ}^{+1.30^\circ } $ and 
$\delta=5.72^\circ \ _{-0.40 ^\circ }^{+0.70^\circ }$.

\begin{figure}[t]
\begin{center}
\includegraphics[width=10.0cm]{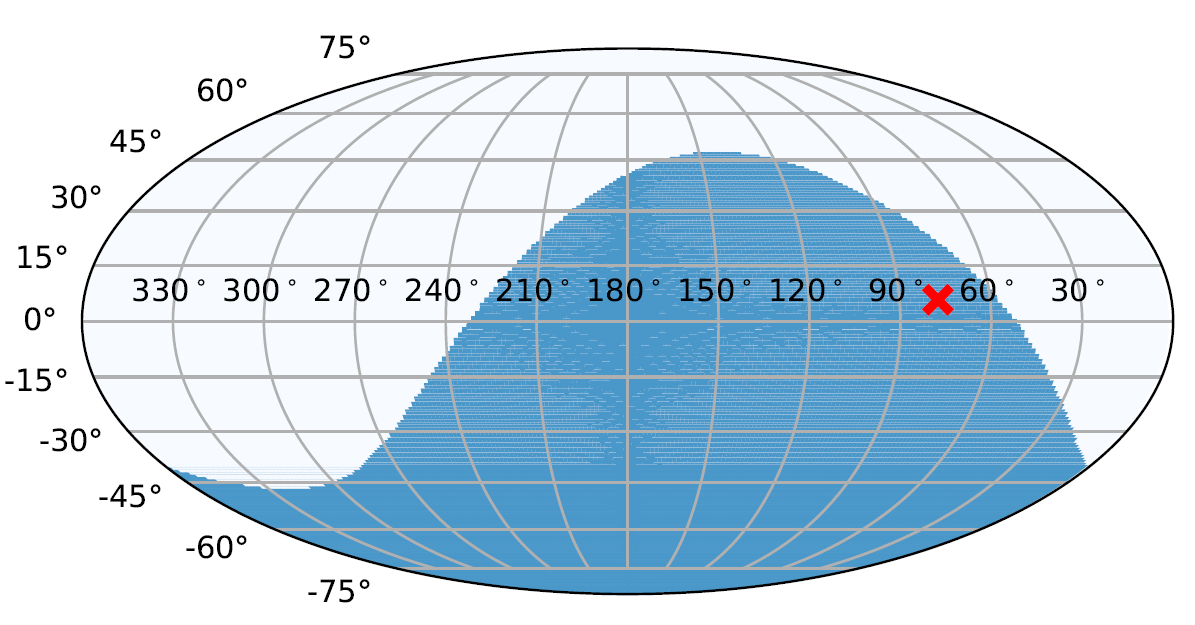}
\end{center}
\caption{\label{fig:txs05-pos} \small
Visibility map for the ANTARES detector of IC170922A (represented by the red marker) in equatorial coordinates. The sky regions below and above the horizon at the alert time are shown in blue and white, respectively. Events that originate from the blue (white) region will be seen as upgoing (downgoing) in the detector frame.}
\end{figure}

The reconstructed direction of IC170922A corresponded, at the location of the ANTARES detector, to a direction $14.2^\circ $ below the horizon, Fig. \ref{fig:txs05-pos}.
A possible neutrino candidate would thus be detected as an upgoing event.

Based on the originally communicated location of IC170922A, HE neutrino candidates were searched for in the ANTARES online data stream, relying on a fast algorithm that selects only upgoing neutrino track candidates \cite{tatoo16}.
This algorithm uses an idealized detector geometry and no information yet about the dynamical positioning calibration.
At 10 TeV, the median angular resolution for muon neutrinos is below 0.5$^\circ$. 
For neutrino energies below $\sim$ 100 TeV, ANTARES has competitive sensitivity to this position in the sky.

As a result of this search, no upgoing muon neutrino candidate event was recorded in a cone of $3^\circ$ centered on the IceCube event coordinates and within a $\pm 1$ h time-window centered on the event time. 
A search over an extended time window of $\pm 1$ day also yielded no detection. Averaged over a day, the source is below the ANTARES horizon with a 46\% fraction. 
The result of this study was reported in \cite{atel}. 

This non-detection in $2$ h was used to provide a preliminary constraint on the neutrino fluence, computed as in \cite{PRD}. A point source of neutrinos, with a power-law spectrum $\frac{dN}{dE}\propto E^{-\gamma}$, has been assumed.
For a flux with $\gamma= 2.0$, the 90\% C.L. fluence upper limit is 15 GeV cm$^{-2}$, integrated over the energy range 3.3 TeV - 3.4 PeV (the range corresponding to 5-95\% of the detectable flux).
For $\gamma=2.5$, the 90\% C.L. fluence upper limit is 34 GeV cm$^{-2}$, integrated in the 450 GeV - 280 TeV energy range. 


\section{Time-integrated search for neutrinos from TXS 0506+056 \label{sec:an2}}
\mau{After the IceCube alert, the Fermi-LAT Collaboration announced that the blazar TXS 0506+056, located at 6 arcmin from the center of the estimated IC170922A direction, showed enhanced gamma-ray activity during the week of the alert (Atel \#10791).}
Then, the MAGIC Collaboration observed this source under good weather conditions and a 5$\sigma$ detection above 100 GeV was achieved after 12 h of observations from September 28 till October 3 (ATel \#10817).
\mau{Other $\gamma$-ray ground detectors, namely the H.E.S.S. (ATel \#10787), HAWC (ATel \#10802) and VERITAS (ATel \#10833) Collaborations, reported no significant detection.}

Considering the potential association between IC170922A and the blazar TXS 0506+56, its location was scrutinized following the ANTARES standard point-source method. 
In this approach, the directions of a predefined list of neutrino source candidates are investigated to look for an excess of events and to determine a flux upper limit in case of a null observation.
The results of this study using 106 pre-selected sources (not including TXS 0506+056) have been recently published \cite{ps}. The data were recorded from January 29, 2007, to December 31, 2015, corresponding to a livetime of 2424 days. 

The neutrino event selection is optimized following a blinding procedure on pseudodata sets randomized in time before performing the analysis.
To find clusters of neutrinos from cosmic sources, a maximum likelihood ratio approach is followed. While atmospheric neutrino events are randomly distributed, neutrinos from point-like sources are expected to accumulate in spatial clusters. 
The likelihood describes the data in terms of signal and background probability density functions that include both shower-like and track-like events. 
The signal likelihood includes a parameterization of the PSF and of the probability density function of the reconstructed energy, which is estimated for each event.

The PSF is defined as the probability density to find a reconstructed event at an angular distance $\Delta \Psi$ around the direction of the source. It depends on the angular resolution of the event sample. The PSFs (for track- and shower-like events) are determined from Monte Carlo simulations of neutrinos with an $E^{-2}$ energy spectrum: about 50\% of the track (shower) events are reconstructed within $0.4^\circ$ ($3^\circ$) from the parent neutrino.

The probability density functions for the energy estimators take into account the different energy spectra of atmospheric and cosmic neutrinos. The spectrum of both is expected to follow a power-law type, $\propto E^{-\gamma}$, but with different values of the spectral indexes: $\gamma\simeq 3.6$ for atmospheric neutrinos \cite{antaes} and $\gamma\simeq 2.0$ for cosmic neutrinos.
Thus, the flux of cosmic neutrinos is expected to exceed the atmospheric neutrino flux above a certain energy threshold, estimated around 100 TeV.

\begin{figure}[tbh]
\begin{center}
\includegraphics[width=13.5cm]{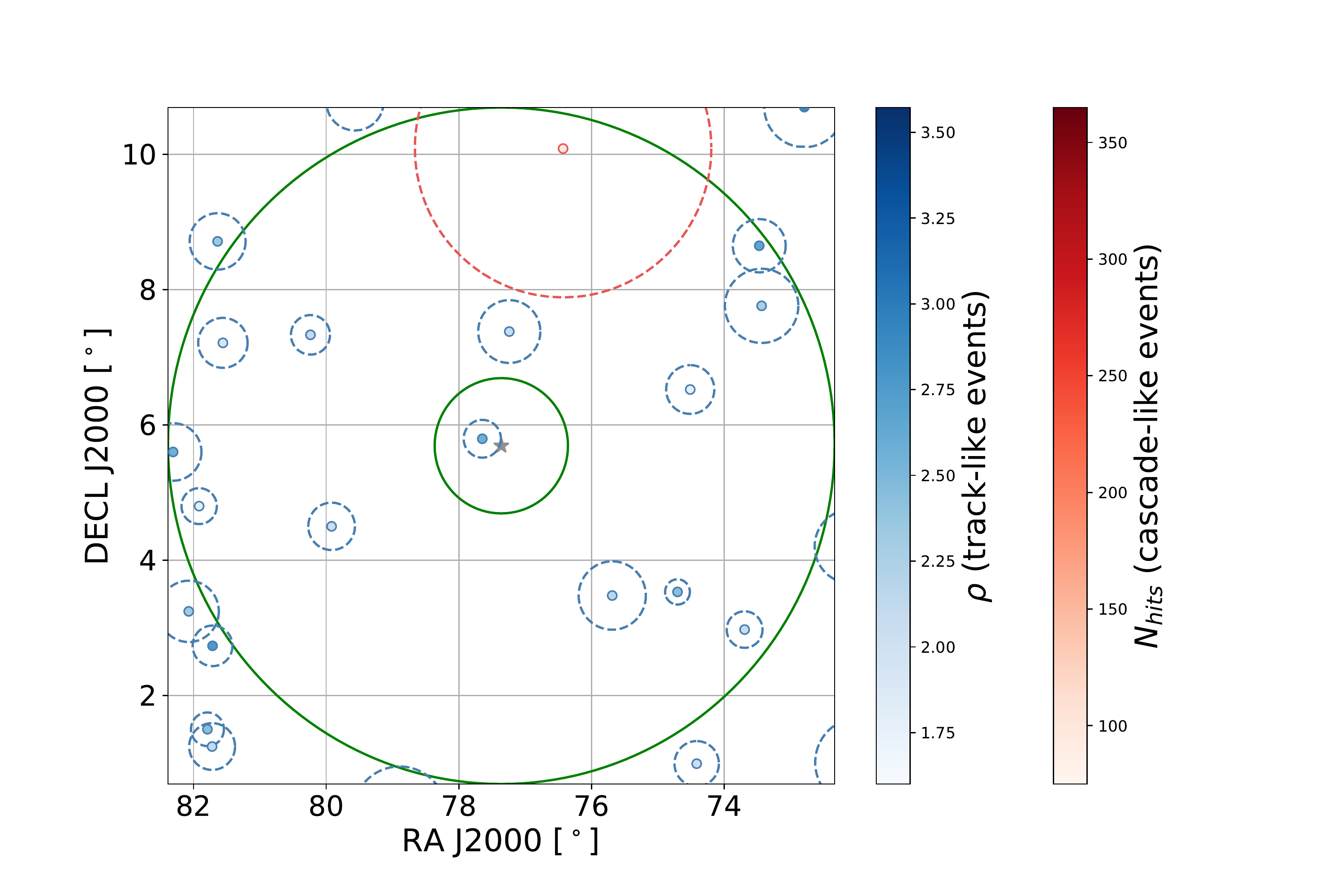}
\end{center}
\caption{\label{fig:event} \small
Distribution of ANTARES events in the (RA, $\delta$) coordinates around the position of TXS 0506+056. The inner (outer) green line depicts the one (five) degree distance from the source position, indicated as a gray star. The red point denotes a shower-like event, whereas the blue points indicate track-like events. The dashed circles around the events indicate the angular error estimate. 
Different shades of red and blue correspond to the values assumed by the energy estimators, the right legend shows the color scales. The number of hits is used for shower-like events and the $\rho$ parameter for track-like events. Refer to \cite{ps} for further details on the energy estimators.}
\end{figure}

The background rate depends on the declination, $\delta$. Given the expected small contribution of a cosmic signal in the overall data set, the background rate is estimated directly from the data. Due to the
Earth's rotation and a sufficiently uniform exposure, the background is considered independent of right ascension.
The expected number of background events at the declination of IC170922A in the livetime of 2424 days is 0.18/deg$^2$ and $4\times 10^{-3}$/deg$^2$ for track-like and shower-like events, respectively.

The blazar TXS 0506+056 was added to the list of 106 already studied objects, and analyzed without changing the pre-defined conditions. The corresponding number of signal events, $\mu_{sig}$, which fits the likelihood signal function for this source is $\mu_{sig}=1.03$.
This signal likelihood of $\mu_{sig}$ corresponds to a pre-trial p-value of 2.6\% to be compatible with the background-only hypothesis.
By referring to the candidate list of Table III of \cite{ps}, only for the directions of two other objects, a smaller p-value had been evaluated: HESSJ0632 + 057, with a pre-trial p-value of 0.16\%, and PKS1440-389, with pre-trial p-value of 0.5\%.  After TXS 0506+056, the fourth position corresponds to PKS0235 + 164, with 5\%.
When considering that 107 sources have been investigated, the post-trial p-value for TXS 0506+056 is 87\%. 

There is one track-like event mostly influencing the fit (see Fig. \ref{fig:event}), whose position is within 1$\sigma$ from the source position. The value of the energy estimator, $\rho$, for this event is such that only 9\% of the neutrino candidates inducing a track have a larger value.  
This event occurred on December 12, 2013 (MJD: 56638.70832).

\begin{table}
{\centering \begin{tabular}{|l|c|c|c|c|c|c|}
\hline 
\# & ($\delta, RA$) & $\Delta \Psi$ & MJD & date & $\rho$ & $f(>\rho)$ \\ 
& deg & deg & & dd/mm/year & (a.u.) & \\ \hline
1 & (5.79, 77.65) & 0.30 & 56638.70832 & 12/12/2013 & 2.58  & 0.09 \\
2 & (7.38, 77.24) & 1.69 & 54601.04530 & 15/05/2008 & 2.10   & 0.40 \\
3 & (3.48, 75.69) & 2.77 & 55396.03988 & 19/07/2010 & 2.16   & 0.33 \\
4 & (4.50, 79.92) & 2.81 & 55585.28089 & 24/01/2011 & 2.05   & 0.47 \\
5 & (6.52, 74.51) & 2.95 & 56143.90394 & 04/08/2012 & 1.78   & 0.87 \\
6 & (7.33, 80.24) & 3.30 & 56268.37325 & 07/12/2012 & 2.11   & 0.39 \\ 
7 & (3.53, 74.70) & 3.42 & 56495.86001 & 22/07/2013 & 2.45   & 0.14 \\
8 &(7.76, 73.44)& 4.41 & 57455.29659 & 08/03/2016 & 2.27 & 0.24 \\ 
9 & (7.21, 81.56) & 4.44 & 56298.41381 & 06/01/2013 & 1.94   & 0.63 \\
10 & (2.97, 73.69) & 4.56 & 54742.73522 & 03/10/2008 & 2.06   & 0.44 \\
11 &(4.80, 81.91) & 4.62 & 57392.49813 & 05/01/2016 & 1.80 & 0.83 \\ 
12& (8.65, 73.47) & 4.86 &  54853.53484 & 22/01/2009 & 2.67  & 0.07 \\
13 & (5.60, 82.31) & 4.92 &  55399.13414 & 22/07/2010 & 2.57  & 0.09 \\ \hline
S1 & (10.08, 76.43) & 4.49 & 55144.74625 & 09/11/2009 & 101 & 0.74 \\ \hline
\end{tabular}\par}
\caption{{\small Neutrino candidates registered by the ANTARES detector within an angular distance $\Delta \Psi$ from TXS 0506+056. There are 13 tracks and one shower event (S1). The table reports for each neutrino candidate: the equatorial coordinates ($\delta, RA$); the angular distance $\Delta \Psi$ from TXS 0506+056; the Modified Julian Date (MJD); the date (dd/mm/year); the energy estimator value (arbitrary units, a.u.); and the fraction, $f$, of events with energy estimator values larger than that of the event. For the shower S1 the energy estimator is the number of hits used in the fit. Refer to \cite{ps} for more details. Events \# 8 and 11 have been recorded during the 2016-2017 period.}
\label{tab:events}}
\end{table}


\mau{Fully calibrated data collected by the ANTARES detector during 2016 and 2017 are available.}
Limited to the TXS 0506+056 source, these additional two years of data have been unblinded using the same predefined conditions used for the 2007-2015 period. The additional livetime corresponds to 712 days.
In this additional sample, there are two new tracks within 5$^\circ$ from the source.

The information concerning the 13 track- and one shower-like neutrino candidates in the period 2007-2017 with coordinates within $5^\circ$ from TXS 0506+056 are reported in Table \ref{tab:events}. Within a radius of $5^\circ$ in a livetime of 3136 days, 17$\pm 4$ atmospheric neutrino events are expected.

After the inclusion of the 2016-2017 data, the number of fitted signal events, $\mu_{sig}$, remains 1.03 and the associated p-value rises from 2.6\% to 3.4\%. 
\mau{Using the total livetime of 3136 days, the corresponding 90\% C.L. upper limits on the flux normalization factor at the energy of 100 TeV, $\Phi_{100\textrm{ TeV}}^{90\%}$, assuming a steady neutrino source and spectral index $\gamma=2.0$, is $1.6\times 10^{-18}$ GeV$^{-1}$ cm$^{-2}$ s$^{-1}$ in the 5\%--95\% energy range 2 TeV--4 PeV.}


\section{Search for neutrinos in the bursting period
\label{sec:an3}}

The time-dependent analysis performed by the IceCube Collaboration \cite{ICalo} contains a significant excess that is identified by two time-window shapes (one Gaussian- and one box-shaped time profile).

The ANTARES Collaboration has developed a time-dependent analysis aimed at finding correlations between neutrinos and high-energy electromagnetic emission, reducing by a factor of 2--3 the signal required for a discovery with respect to a time-integrated search \cite{ANXrays}. 
For TXS 0506+056, a search similar to that reported in \cite{ANTbl} was performed, with a bursting period as defined by the two profiles provided by the IceCube Collaboration, instead of an electromagnetic light curve.

The first one (denoted in the following \textit{Gaussian flare}) is modeled by a Gaussian signal centered on MJD 57004 and with standard deviation $\sigma=55.0$ days. We considered a period $\pm 5\sigma$ wide, corresponding to 550 flaring days.
The second one (referred to as \textit{box flare}) assumes a box-shaped flare starting at MJD 56937.81 and ending at MJD 57096.21, corresponding to 158.40 flaring days. 

The time-dependent analysis first defines the values of selection cuts that, assuming a power-law neutrino spectrum for the signal, optimizes the Model Rejection Factor (MRF). 
Three different spectral indexes for the neutrino spectrum have been considered ($\gamma=2.0, 2.1$ and 2.2), and the cut parameters to be optimized are the maximum reconstructed neutrino zenith angle, $\theta$, and the quality of reconstructed tracks, $\Lambda$.
The resulting optimized set of cuts for the three considered spectral indexes are $\cos\theta>-0.15$ and $\Lambda >-5.6$.
For comparison, the corresponding cuts for the time-integrated search reported in Sect. \ref{sec:an2} are $\cos\theta>-0.10$ and $\Lambda>-5.2$. This means that the time-dependent analysis allows for more downward-going neutrino candidates and tracks with a lower reconstruction quality parameter, refer to Fig. 2 of \cite{ps}.
According to this set of cuts, the expected background during the box flare has been estimated as 0.04 (4) events within 0.5$^\circ$ (5$^\circ$) from the source. This is a factor $\sim$ 4 larger than that determined for the time-integrated search. 
\mau{The background increases because a larger fraction of atmospheric muons is mis-reconstructed and because more low-energy atmospheric neutrinos are included. Thanks to the reduced observational time window, the method allows the presence of lower energy cosmic neutrinos and, \textit{de facto}, corresponds to an analysis sensitive to a softer neutrino energy spectrum.}

After the optimization, the data have been unblinded. 
The results are compatible with the expectation from the atmospheric background, and no signal has been found during neither of the considered flares. 
Within 2$^\circ$ from the source, 10 background events are expected during the analyzed period while 13 events have been found in data. None of them lies within either of the two considered flaring periods. 
From these null results, 90\% C.L. upper limits have been derived for the neutrino flux. 
\mau{For an $E^{-2.0}$ ($E^{-2.1}$) [$E^{-2.2}$] spectrum they correspond to a normalization factor of 
$\Phi_{100\textrm{ TeV}}^{90\%}= 4.6 (4.4) [4.2] \times 10^{-18}$ GeV $^{-1}$ cm$^{-2}$ s$^{-1}$ 
for the Gaussian-shaped period. 
The energy range containing the 5-95\% of the detectable flux is 2.0 (1.3) [1.0] TeV -- 3.2 (1.6) [1.0] PeV.
For the box-shaped period, the flux normalization factors are a factor 3.3 higher.}



\section{Conclusions\label{sec:an4}}

The GCN circular \cite{ICGCN} on the EHE neutrino candidate event, IC170922A, has promptly triggered a search for neutrino candidates in the ANTARES online data stream. The position of the IC170922A event at the location of the ANTARES detector is 14.2$^\circ$ below the horizon. 
For neutrino energies below $\sim$ 100 TeV, ANTARES has competitive sensitivity with respect to IceCube to this position in the sky.
No upgoing muon neutrino candidate event was recorded within 3$^\circ$ around the IC170922A direction within $\pm 1$ h centered on the event time.

Successively, due to the potential association between IC170922A and the blazar TXS 0506+56, this source has been scrutinized following the ANTARES standard point-source method. 
TXS 0506+056 was not considered among the 106 pre-selected sources in our recent study \cite{ps}, and a dedicated analysis was done using two additional years of data. The considered period (from 2007 to 2017) corresponds to a total livetime of 3136 days. 
The result yields a number of fitted signal events $\mu_{sig}= 1.03$ with an associated pre-trial p-value of 3.4\%.
This is the third most significant source of our list. When considering that 107 sources have been investigated, the post-trial p-value for TXS 0506+056 corresponds to 87\%. 

Finally, according to the results of a time-dependent analysis performed by the IceCube Collaboration \cite{ICalo}, we applied our standard time-dependent analysis on a bursting period centered on MJD 57004 (December 13, 2014).
The analysis, which relies on relaxed selection criteria on the neutrino direction and reconstructed track quality, yielded no events within the Gaussian and the box-shaped periods defined by the IceCube analysis.

\section*{Acknowledgement} 
{\small We thank the IceCube Collaboration for sharing information before the public release. 
The authors acknowledge the financial support of the funding agencies:
Centre National de la Recherche Scientifique (CNRS), Commissariat \`a
l'\'ener\-gie atomique et aux \'energies alternatives (CEA),
Commission Europ\'eenne (FEDER fund and Marie Curie Program),
Institut Universitaire de France (IUF), IdEx program and UnivEarthS
Labex program at Sorbonne Paris Cit\'e (ANR-10-LABX-0023 and
ANR-11-IDEX-0005-02), Labex OCEVU (ANR-11-LABX-0060) and the
A*MIDEX project (ANR-11-IDEX-0001-02),
R\'egion \^Ile-de-France (DIM-ACAV), R\'egion
Alsace (contrat CPER), R\'egion Provence-Alpes-C\^ote d'Azur,
D\'e\-par\-tement du Var and Ville de La
Seyne-sur-Mer, France;
Bundesministerium f\"ur Bildung und Forschung
(BMBF), Germany; 
Istituto Nazionale di Fisica Nucleare (INFN), Italy;
Nederlandse organisatie voor Wetenschappelijk Onderzoek (NWO), the Netherlands;
Council of the President of the Russian Federation for young
scientists and leading scientific schools supporting grants, Russia;
National Authority for Scientific Research (ANCS), Romania;
Mi\-nis\-te\-rio de Econom\'{\i}a y Competitividad (MINECO):
Plan Estatal de Investigaci\'{o}n (refs. FPA2015-65150-C3-1-P, -2-P and -3-P, (MINECO/FEDER)), Severo Ochoa Centre of Excellence and MultiDark Consolider (MINECO), and Prometeo and Grisol\'{i}a programs (Generalitat
Valenciana), Spain; 
Ministry of Higher Education, Scientific Research and Professional Training, Morocco.
We also acknowledge the technical support of Ifremer, AIM and Foselev Marine
for the sea operation and the CC-IN2P3 for the computing facilities.
 }


\end{document}